\documentclass[11pt,aps,onecolumn]{revtex4}   

\usepackage{amsmath}    
\usepackage{graphicx}   

\begin{document}

\title{Electrodynamics of Perfect Conductors}
\author{Miguel C. N. Fiolhais}
\email{miguel.fiolhais@cern.ch}   
\affiliation{LIP, Department of Physics, University of Coimbra, 3004-516 Coimbra, Portugal}
\author{Hanno Ess\'en}
\email{hanno@mech.kth.se}   
\affiliation{Department of Mechanics, KTH, 10044 Stockholm, Sweden}

\date{\today}

\begin{abstract}
The most general electrodynamic equations of a perfect conducting state are obtained using a variational principle in a classical framework, 
following an approach by Pierre-Gilles de Gennes.
London equations are derived as the time-independent case of these equations, corresponding to the magnetostatic minimal energy state of the perfect conducting system.
For further confirmation, the same equations are also derived in the classical limit of the Coleman-Weinberg model, the most successful quantum macroscopic theory of superconductivity.
The magnetic field expulsion is, therefore, a direct consequence of zero resistivity and not an exclusive property of superconductors.
\\
\\
\noindent The following paper is published in the International Journal of Theoretical Physics: {http://link.springer.com/article/10.1007\%2Fs10773-013-1491-9}

\end{abstract}

\maketitle

\section{Introduction}
\label{intro}
With the discovery of the magnetic field expulsion from superconductors by Meissner and Ochsenfeld, in 1933 \cite{meissner}, 
Meissner claimed the effect had no classical explanation, creating a distinction between perfect conductors and superconductors.
Even though this statement has been repeatedly contested and disproved in the scientific literature throughout the years, the magnetic field expulsion of a superconductor is still presented in most textbooks as a pure quantum effect. 
This view has been refuted by
Cullwick \cite{cullwick}, de Gennes \cite{degennes}, Pfleiderer \cite{pfleiderer}, Karlsson \cite{karlsson}, Bad\'ia-Maj\'os \cite{badiamajos}, Kudinov \cite{kudinov}, Mahajan \cite{mahajan}, Fiolhais \emph{et al.} \cite{fiolhais}, among others, by reaching the conclusion that the magnetic flux expulsion from a superconductor corresponds to an approach to the state of minimum magnetic energy.
A recent review by Ess\'en \emph{et al.} \cite{essen} addresses the arguments that lead Meissner to his conclusion and to decades of misunderstandings.

One hundred years after its discovery, superconductivity is far from being a well understood phenomenon neither at the microscopic nor at the macroscopic level.
This letter attempts to provide further clarification on the macroscopic behavior of superconductivity, in particular the Meissner effect, by establishing a bridge with perfect conductivity.
The most general equations of a perfect conducting system are derived by minimizing the action of the electromagnetic field and the magnetic field expulsion, \emph{i.e.} the London equations, 
corresponds to the magnetostatic energy minimum state (equilibrium solution). In this letter we also show that this result is corroborated and validated by the classical limit of the most successful quantum field theory of superconductivity built to date, the Coleman-Weinberg model.

It must be stressed that superconductivity is far from being a classical phenomenon, an idea not supported by the authors nor that superconductors and perfect conductors are the same thing. 
The magnetic field flux expulsion appears in the classical limit of a quantum theory of superconductivity, as a result of perfect conductivity, \emph{i.e.} in the absence of dissipative losses.
Consequently, Meissner effect does not distinguish superconductivity from perfect conductivity as it is often stated, instead, the main distinction comes from the phase transition. 

\section{Classical Electrodynamics Framework}

A classical derivation of the London equations for the magnetic field expulsion based on a minimum 
energy principle can be found in de Gennes textbook \emph{Superconductivity of Metals and Alloys} \cite{degennes}.
The internal energy density of a perfect conductor is assumed to have two contributions, namely 
the electromagnetic field energy density $\mathcal{U}_{EM}$ and the kinetic energy density of the moving superconducting charge carriers $\mathcal{U}_K$:
\begin{eqnarray}
\mathcal{U} &=&  \mathcal{U}_{EM} + \mathcal{U}_K \nonumber \\
\mathcal{U} &=&  \frac{1}{2} \mathbf{B} ^2 + \frac{1}{2} n m \mathbf{v}^2 \, ,
\end{eqnarray}
where $n$ is the density of charge carriers of mass $m$ and charge $e$. 
The magnetic field energy is considered 
to be separate from the kinetic energy of the conduction electrons, which is valid in the classical approach, and the drift velocity of the free electrons is assumed to be approximately equal to 
their total velocity, true in the low temperature limit.
By means of the Maxwell equation $ \nabla \times \mathbf{B} = \frac{1}{c} \mathbf{j}$ and $\mathbf{j}=en \mathbf{v}$, the total energy of the system becomes:
\begin{eqnarray}
U &=&  \int \mathcal{U}_{EM} + \mathcal{U}_K \,  \textrm{d} V \nonumber \\
U &=& \frac{1}{2}  \int   \mathbf{B} ^2 + \lambda_{L}^2 \left ( \nabla \times  \mathbf{B} \right )^2 \,  \textrm{d} V \, ,
\end{eqnarray}
where $n$ is assumed to be constant in the region where there is current, and the London penetration depth is,
\begin{eqnarray}
 \lambda_{L} = \sqrt{\frac{mc^2}{n e^2}} \, .
\end{eqnarray}
By minimizing the energy with respect to the magnetic field, the London equation appears naturally,
\begin{eqnarray}
\mathbf{B}  -  \lambda_{L}^2 \nabla^2 \mathbf{B} = 0.
\end{eqnarray}
The conclusion is summarized in de Gennes book as: \emph{``The superconductor finds an equilibrium state where the sum of the kinetic and magnetic energies is minimum, and this state, for macroscopic samples, corresponds to the expulsion of magnetic flux''}.

Here we further extend the work of de Gennes by considering the action for the perfect conducting system and applying the variational principle to this action which allows us to derive most general dynamic equations of the electromagnetic field inside a perfect conductor. In this case, unlike the static case, the electric field appears, induced by the variation of the magnetic field (Faraday-Lenz law). Note that the perfect conductor is electrically neutral at each point, $i.e.$ the charge density $\rho$ is zero everywhere, therefore, no Coulomb potential is considered.
By including the contribution from the induced electric field, and by using the Maxwell equation $\nabla \times \mathbf{B} = \frac{1}{c} \mathbf{j} + \frac{1}{c} \frac{ \partial \textbf{E}}{\partial t }$, the total energy density of the system becomes,
\begin{eqnarray}
\mathcal{U} &=&  \frac{1}{2} \mathbf{E} ^2 +  \frac{1}{2} \mathbf{B}^2 + \frac{1}{2} n m \mathbf{v}^2 \nonumber \\ 
\mathcal{U} &=&  \frac{1}{2} \mathbf{E} ^2 +  \frac{1}{2} \mathbf{B}^2 + \frac{1}{2} \frac{m}{n e^2} \mathbf{j}^2 \nonumber \\ 
\mathcal{U} &=&  \frac{1}{2} \mathbf{E} ^2 +  \frac{1}{2} \mathbf{B}^2 + \frac{\lambda_{L}^2}{2} (\nabla \times \mathbf{B})^2 \nonumber \\ 
&+& \frac{\lambda_{L}^2}{2} \frac{1}{c^2} \left(\frac{\partial \mathbf{E}}{\partial t}\right)^2 -  \frac{\lambda_{L}^2}{c} \frac{\partial \mathbf{E}}{\partial t} \cdot (\nabla \times \mathbf{B}).
\end{eqnarray}
Therefore, the full lagrangian density of the perfect conducting system is:
\begin{eqnarray}
\mathcal{L} &=&  F_{\mu 0} \frac{ \partial A^\mu } { \partial t } - \mathcal{U} \nonumber \\ 
\mathcal{L} &=&  \frac{1}{2} \mathbf{E} ^2 -  \frac{1}{2} \mathbf{B}^2 -  \frac{\lambda_{L}^2}{2} (\nabla \times \mathbf{B})^2 \nonumber \\
                     & - &  \frac{\lambda_{L}^2}{2} \frac{1}{c^2} \left(\frac{\partial \mathbf{E}}{\partial t}\right)^2  + \frac{\lambda_{L}^2}{c} \frac{\partial \mathbf{E}}{\partial t} \cdot (\nabla \times \mathbf{B}).
\end{eqnarray}
Applying the usual variational principle to the action constructed from this Lagrangian, the Euler-Lagrange equations for $\mathbf{E}$ and $\mathbf{B}$ are readily obtained,
\begin{equation}
\label{kleinB}
\mathbf{B}  -   \lambda_{L}^2 \nabla^2 \mathbf{B} + \frac{\lambda_{L}^2}{c^2} \frac { \partial^2 \mathbf{B} }{ \partial t^2 }  =  0,
\end{equation}
and,
\begin{equation}
\label{kleinB2}
\mathbf{E}  -   \lambda_{L}^2 \nabla^2 \mathbf{E} + \frac{\lambda_{L}^2}{c^2} \frac { \partial^2 \mathbf{E} }{ \partial t^2 } =  0.
\end{equation}
These equations are Klein-Gordon-like equations for the components of the electric and magnetic fields, which implies the electromagnetic field must be highly suppressed inside the perfect conductor.
In the stationary regime, the equation for the electric field is fulfilled with $\mathbf{E} = 0$, and the equation for the magnetic field reduces to the London equation of the magnetic field expulsion out of a superconductor, namely: $\mathbf{B}  -  \lambda_{L}^2 \nabla^2 \mathbf{B} = 0$. In summary, the London equation is the time-independent variational equation derived from the action of the electromagnetic field inside a perfect conductor.

\section{Coleman-Weinberg Model}

Soon after the discovery of the Meissner effect, London brothers developed the first phenomenological description of the electromagnetic field in a superconductor \cite{london}. Later on, in 1950, Ginzburg and Landau \cite{ginzburg} proposed a macroscopic quantum theory able to describe the phase transition and predicting the existence of a coherence length in superconductors.
The ratio between the London penetration length and the coherence length, \emph{i.e.} the Ginzburg-Landau parameter $\kappa$, determines the dividing line between type-I and type-II superconductors at $\kappa=1/\sqrt{2}$ \cite{halperin}. Moreover, the tricritical point, which separates first and second order phase transitions in superconductors, was predicted to lie near the type-I and type-II dividing line at $\kappa \approx 0.81/\sqrt{2}$ \cite{kleinert,kleinert2,kleinert3}, and later confirmed by Monte Carlo simulations to lie 
at $\kappa=0.76/\sqrt{2} \pm 0.04$ \cite{hove}.
The first microscopic theory of superconductivity (BCS theory) was established in 1957 by Bardeen, Cooper and Schrieffer \cite{cooper,bardeen,bardeen2} and, also in the same year, Abrikosov predicted the penetration of strong magnetic fields in type-II superconductors through quantum vortices \cite{abrikosov}, giving further credibility to the Ginzburg-Landau model. The first quantum field theory of superconductivity came in 1973 by Coleman and Weinberg \cite{coleman}, as a four-dimensional version of the Ginzburg-Landau model. The Coleman-Weinberg model is a scalar field theory with a quartic interaction, similar to the Higgs model, proposed in 1964, to explain how particles acquire mass in the Standard Model of particle physics through the spontaneous symmetry breaking mechanism \cite{anderson,englert,higgs,guralnik}.

The classical Lagrangian of the Coleman-Weinberg model reads
\begin{eqnarray}
\mathcal{L} & = & \frac{1}{2} \left (  \partial^\mu + i e A^\mu \right ) \phi^* \left (  \partial_\mu - i e A_\mu \right ) \phi \nonumber \\
&& - \frac{1}{2} \mu^2 |\phi|^2 - \frac{\lambda}{4} |\phi|^4 - \frac{1}{4} F^{\mu\nu}F_{\mu\nu} \, ,
\end{eqnarray}
where $e$ represents the electric charge, and 
$\lambda$ gives the strength of the quartic term.
The scalar (spin-0) field is represented by $\phi$ with a mass term $\mu^2$, and $A^\mu$ is the electromagnetic four-potential with $F^{\mu\nu} = \partial^\mu A^\nu - \partial^\nu A^\mu$. 
The ground state of the potential, $V(\phi) = \frac{1}{2} \mu^2 |\phi|^2 + \frac{\lambda}{4} |\phi|^4 $, is particularly interesting if $\mu^2 < 0$.
The minimum of the potential is obtained for an infinite number of degenerate states satisfying:
\begin{equation}
 \phi  ^2 = - \frac{\mu^2}{\lambda} = \frac{|\mu^2|}{\lambda} = \nu^2 \, ,
\end{equation}
where $\nu$ represents the vacuum expectation value.
The complex scalar field $\phi(x)$ can be parametrized around the minimum in the most general form in terms of two real fields $h(x)$ and $\theta(x)$,
\begin{equation}
\phi (x) = \left ( h(x) + \nu \right ) e^{i\theta(x)/\nu} \, .
\end{equation}
By taking a particular gauge choice, $\theta (x) = 0$, the symmetry is spontaneously broken, and the Lagrangian becomes,
\begin{eqnarray}
\mathcal{L} & = & \frac{1}{2} \left ( \partial^\mu  h  \right ) ^2 -\lambda \nu^2 h^2 + \frac{1}{2} e^2\nu^2 A_\mu^2 - \lambda \nu h^3 - \frac{1}{4} \lambda h^4  \nonumber \\
&&  + \frac{1}{2} e^2h^2 A_\mu^2 + \nu e^2 A_\mu^2 h - \frac{1}{4} F^{\mu\nu}F_{\mu\nu} \, .
\end{eqnarray}
where the vector gauge boson field $A^\mu$ appears with a mass term, $m_A = e \nu$, known as the Meissner-Higgs mass \cite{kleinert2}. The Euler-Lagrange equation for the gauge field $A_\mu$ yields,
\begin{eqnarray}
\label{eqA}
\partial_\mu F^{\mu\nu} = j^\nu  & = &   - e^2 \left ( h + \nu \right )^2  A^\nu  \nonumber  \\ 
 & = &   - e^2 \phi ^2  A^\nu \, ,
\end{eqnarray}
which is equivalent to the London equations, where $\phi^2$ is proportional to the charge carriers density in the superconductor. 
It is worth noting, however, that the existence of a quartic interaction in the superconductor, which generates a vacuum 
expectation value different from zero and explains the phase transition, naturally increases the value of $\phi^2$ and enhances the strength of Meissner effect as in comparison with a perfect conductor (normal QED). 

In conclusion, the London equations were derived using the classical Lagrangian of the Coleman-Weinberg model, \emph{i.e.} without applying any quantum corrections.
Thus, the magnetic field expulsion from a superconductor shall not be regarded as a genuine quantum effect.

\section{Summary}

There are two main conclusions to draw from this letter: 1) magnetic field expulsion, the \mbox{Meissner} effect, is not a quantum effect - the London equation can be derived classically or obtained as a classical limit of a quantum theory, and 2) the Meissner effect is not restricted to superconductors only, appearing as a consequence of perfect conductivity, independently of the mechanism that leads to zero resistance. 
These conclusions were drawn as a direct result for a perfect conductive system in a classical framework and as a classical limit of a quantum field theory, the Coleman-Weinberg model.

\end{document}